\documentclass[11pt]{article}

\usepackage[hang,small,labelfont=bf,up,textfont=it,up]{caption}
\usepackage{booktabs}
\usepackage{float}
\usepackage{hyperref}
\usepackage{abstract}
\usepackage{titlesec}
\usepackage[left=2.5cm,right=2.5cm,top=2.5cm,bottom=2.5cm]{geometry}
\usepackage{graphicx}
\usepackage{amsmath}
\usepackage{mathrsfs}
\usepackage{amsfonts}
\usepackage{amsthm}
\usepackage{amssymb}
\usepackage{braket}
\usepackage{array}
\usepackage{cancel}

\title{\fontsize{15pt}{5pt}\selectfont\textbf {
Addressing the so called quantum/classical ``divide'' in gravitational contexts, and its implications in cosmology}}
\author{ {\large Rosa-Laura Lechuga-Solis \& Daniel Sudarsky }\\
{\normalsize Instituto de Ciencias Nucleares, UNAM A. Postal 70-543, M\'exico City 04510, M\'exico} \\
{\normalsize rlechuga@correo.nucleares.unam.mx, sudarsky@nucleares.unam.mx}
}

\date{}

\begin{document}

\maketitle

\begin{abstract}
\noindent 

 The manner in which one approaches the interface between gravitation and quantum theory is influenced by one's posture regarding quantum mechanics and the issues that revolve about its interpretational problems. We discuss here the way in which these issues occur in the inflationary cosmology setting, the serious confusions that ensue, and one path that we have used to deal with the problem which perhaps surprisingly ends up, not just offering a conceptually clear physical picture, but also modifying some of the standard observational predictions of the inflationary approach.

\end{abstract}

\section{Introduction}

Besides his remarkable work on cosmology, Georges Lemaître was very interested in quantum theory, and even in considering both topics together \cite{Lemaitre}. It seems thus appropriate to honor his memory by discussing issues at the interface of the two subjects. 

It is quite common to hear discussions about the {\it quantum-classical transition}, particularly in cosmological contexts. Let us start by clarifying that we hold the view that, at the fundamental level, all aspects of nature require a quantum description, and, thus, that there is not, strictly speaking, any such thing as a classical-quantum divide. There are, however, in most situations that we are able to treat with our physical theories, some aspects that admit a classical treatment. Thus one can discuss the issue by framing it as the question of when and under which conditions can a classical description of the physical situation at hand be considered sufficiently ``good'' for the intended purposes and the sought for precision. 

When considering gravitation we, of course, face the problem that, at this point, we have no fully developed and workable theory of quantum gravity. This fact is exemplified by our inability to produce anything close to a sensible mathematical characterization of, say, the space-time corresponding to the Moon in a state of quantum superposition of being between the Earth and the Sun and the situation where the Earth and te Moon's places are exchanged \footnote{We might, of course, be naturally inclined to argue that such situations  simply do {\it not occur in nature}, but then we should be able to point out at the features in our theories that account for that. Many cosmologists believe that this is, in fact, explained by decoherence, but a careful look at the matter shows the fallacy of those beliefs (see for instance \cite{D-E1} and \cite{decoh2}).}. Of course, this by no means is to be taken as dismissive of the enormous efforts made by colleagues working of the various existing approaches to the subject, but perhaps should be taken as indicative of the extreme difficulty of the enterprise. 
\newpage 
{\bf A word about methodology:}

We approach the exploration of the GR/ QT regime in what we call a top-bottom approach. The usual bottom-up approach starts by postulating a fundamental theory (String Theory \cite{strig1}, Loop Quantum Gravity\cite{lqg2}, Causal sets\cite{cs1}, Causal Dynamical Triangulations\cite{dt2}, etc.) and devote the work to attempts to connect such proposal to the regimes of interest of the "world out there": Cosmology, Black Holes, etc. The top-bottom approach, in contrast, aims to push existing, well-tested, and developed theories, to address open issues that seem to lie beyond their domain. Possible modifications of such theories might serve as clues about the nature of the more fundamental theory. 
 
\medskip 

The idea is, therefore, to use General Relativity together with Quantum Field Theory (the curved space-time version), that is semi-classical gravity, while considering at the same time the foundational difficulties that afflict quantum theory in general. In particular, we will be concerned about the issue known as the ``measurement problem" \cite{MP}, the \textbf{M problem}, which we briefly discuss in appendix A, since it is, unfortunately, an issue often ignored in discussions about cosmology.

\vspace{0.3cm}

\section{The central issue}

One should be worried about any analysis of a physical situation in which the theoretical framework employed is not clearly and precisely stated, particularly when there are various possible choices, as there is a danger of relying on a changing framework within the discussions, which can easily lead to self-contradictory conclusions. 

Our concern here is specifically with how quantum uncertainties are treated and, more importantly, how they are interpreted. {\bf Quantum uncertainties or indefiniteness} are often referred to as quantum fluctuations. This terminology is rather unfortunate as it promotes confusion with stochastic fluctuations, which in turn, can refer to either, small changes in a system occurring in time, or to variations in localized aspects of an extended system, or among individuals within an actual ensemble.  
 \\
 
The above notions and the concomitant confusions appear in various places in the discussions of our cosmological ideas and theories:
\begin{itemize}
 \item[i)] The emergence of the seeds of cosmic structure during inflation.
 \item[ii)] The generation of primordial gravity waves during inflation.
 \item[iii)] Eternal Inflation.
\end{itemize}
 
 
 \medskip
 
 
Quantum uncertainties might justifiably be taken as measures of a {\it ``stochastic''} kind of variation, in connection with an appropriate measurement, thus bringing in the ``$M$ problem". However, it must be emphasized that, at least within the textbook accounts of quantum mechanics, in the absence of a measurement-like process, those quantum uncertainties cannot be interpreted as actual stochastic fluctuations of any kind. One, of course, might want to adopt a different posture regarding quantum theory and the way it views quantum uncertainties, but then one should be explicit and clear about what that posture is and should use it consistently throughout the discussion. 
 
\smallskip
 
We should, in fact, note that the cosmological context is such that no satisfactory candidate for either observer or measuring device can be identified. As we will see, taking the view that we, as current explorers of the universe, somehow play the role of the external observers bringing about measurement-like processes on the universe that would justify taking the quantum uncertainties as stochastic fluctuations, would, of course, lead us to conceptually problematic notions of cause and effect, as in that case, we would conclude that we played a central role in the generation of primordial cosmic inhomogeneities, which, of course, are a prerequisite for the formation of galaxies, stars, planets, and for life itself to arise. 

\medskip

Simply overlooking the conceptual issues and adopting a ``practical posture" where one thinks that a quantum system just ``jumps around" within the corresponding uncertainty range, is untenable on various grounds. Indeed, such practical posture (``shut up and compute") would bring us into serious conflict with actual experiments:
 
\medskip
 
As rather a clear example of a situation where this type of confusion would lead us astray, we can point to the searches for a non-vanishing Electric Dipole Moment of the Neutron, 
where the quantum uncertainties of order $\sim 10^{-14} \text{e-cm}$ would, 
if viewed in that way, be incompatible with the attainment of the current bounds, $ d_n<10^{-26} \text{e-cm}$, which, must be emphasized, are {\it over 10 orders of magnitude tighter than the corresponding quantum uncertainty!} (these results are obtained in experiments that can best be described in terms of what is known as  {\it ``weak measurements"} as shown in \cite{EDM}).

\section{Cosmic Inflation} 

Contemporary cosmology includes inflation as one of its most attractive components. It represents a very early stage of accelerated expansion (close to de-Sitter), which ``turns off" after at least $N=60$ or so e-folds. This drives the universe into an extremely flat, homogeneous, and isotropic stage, where all matter content (and especially all defects remnant from earlier phase transitions) are exponentially diluted making them irrelevant \, and which solves other ``naturalness problems" \cite{Naturalness, i2}. 
 
The simplest version of the inflationary theory is described by the action,

\begin{equation}\label{GR-Infl}
 S = \int d^{4}x \sqrt{-g} \left\{ \frac{R}{16 \pi G} - \frac{1}{2} \left( \nabla^a \phi \nabla_a \phi + V(\phi) \right) \right\}.
\end{equation}

The inflationary behavior is the result of a scalar field whose potential acts as a slowly varying cosmological constant $V(\phi(t))$. However, its biggest success goes beyond the naturalness issues and is considered to be ``the account" for the emergence of the seeds of cosmic structure as a result of ``quantum fluctuations" with the correct spectrum\cite{f1, f2, f3, f4}. However, at the theoretical/conceptual level, that account is not truly satisfactory, as it relies on serious misconceptions about quantum theory, and on unjustifiable identifications of concepts that are fundamentally different. 

 As already noted the ``quantum fluctuations"  can not simply be taken as if they were  ``stochastic fluctuations" which one might look upon as representing space-time variations and were thus connected to seeds of cosmic structure.

\subsection{Standard treatment for the primordial cosmic structure}

Let's see this in detail (see for instance \cite{f5}). Recall that the starting point of the analysis is considered perturbations for both, the inflaton field $\phi=\phi_{0}+\delta \phi(\eta,\vec{x})$, and the metric which (using the Newtonian gauge) have the form:

\begin{equation}\label{line-element}
ds^{2}=a^{2}(\eta)\left\{ -(1+2\Psi)d\eta^{2}+ \left[(1-2\Psi)\delta_{ij} + \delta h_{ij }\right]dx^{i}dx^{j} \right\} 
\end{equation}

where, $\delta\phi, \Psi, \delta h_{ij}$ are ``small perturbations" containing the spatial dependencies, which might or might not be present. 

The ``background" $ (a(\eta), \phi_0)$ is treated classically and assumed to be dominated by the inflaton potential. The setting moreover includes the assumption of a ``slow roll regime" so that $\phi_0$ changes slowly, and $a(\eta)$ is approximately given by $a(\eta)=\frac{-1}{\eta H_I}$.
It is convenient to set the present value of the scale factor $a$, as $1$, and to denote the period when inflation takes place as corresponding to $\eta \in (-{\cal T}, \eta_0)$ and $\eta_0<0$. 

The ``perturbations": $\delta\phi, \Psi , \delta h_{ij}$, are treated quantum mechanically and they are assumed to be characterized by a Bunch-Davies or adiabatic ``vacuum state" $|0 \rangle$. The state of the quantum field (as any quantum state) is {\it ``also characterized"} by the so-called ``quantum fluctuations".

In the usual treatments, those quantum indeterminacies are {\bf unjustifiably} identified as the primordial inhomogeneities which eventually evolved into all the structure in our Universe: galaxies, stars, planets, etc... and left their traces in the CMB. Here, we face an instance of the kind of confusion discussed before, the identification of quantum uncertainties with stochastic fluctuations.  

To start, we note that according to the inflationary picture: The Universe was Homogeneous and Isotropic (H\&I), (both in the aspects that are described in ``classical", and ``quantum" terms) as a result of inflation \footnote{This characterization is thought to be adequate up to exponentially small corrections characterized by the number of e-folds i.e. $e^{-N}$. The homogeneity of the state can be seen by applying to it a displacement operator given by $\vec D$ on it: $e^{i\hat{\vec P}\cdot \vec{D}|0\rangle=|0\rangle}$, where $\hat{P}$ is the momentum operator.}  The quantum fluctuations encoded in the state of the quantum field do not take away the homogeneity and isotropy of the adiabatic vacuum state (just as in the case of the flat space-time Minkowski vacuum, which is also completely homogeneous and isotropic).

The ``present" universe with galaxies, stars, and planets, including living creatures like us, is clearly not homogeneous and isotropic. Indeed, departures from those symmetries are a precondition for the existence of life! How does this transmutation occur when the dynamics of the closed system does not break those symmetries? 
This issue represents a particularly clear instance of the \textbf{M problem} we have mentioned before.

We  should mention that various  attempts to justify   the problematic  steps  have been  put forward
\cite{dec1, dec2, dec3}  which  have however  been shown to  be   unsatisfactory\cite{s, s2, ss2}.

\section {Collapse theories} 
 
In order to be very clear about the motivation for introducing these theories, one should start by focusing on the measurement problem. We do not want to discuss that subject in detail here, although a very transparent illustration will be presented in Appendix A for the benefit of readers unfamiliar with the corresponding literature. 

A very useful way to frame the \textbf{M problem} is given by \cite{Tim-trilema}, which demonstrates that the following three premises cannot be held simultaneously in a self-consistent manner, 
\begin{itemize}
 \item[P1)] {\bf The characterization of a system by its wave function is complete.}
 Its negation leads, for instance, to hidden variable theories, for example, Bohmian Mechanics.
 \item[P2)] {\bf The evolution of the wave function is always according to Schr\"odinger's equation.} Its negation leads, for instance, to \textit{Spontaneous collapse theories}.
 \item[P3)] {\bf The results of experiments lead to definite results.}
 Its negation leads, for instance, to Many World/ Minds Interpretations, Consistent Histories approach, etc. 
\end{itemize}
 
Collapse theories emerge as a proposal to address the measurement problem by providing 
 a unified version of the two kinds of evolution laws, the unitary $U$ process, and the reduction $R$ process \cite{UyR}.
There is a large amount of previous work on the matter: including the GRW theory (from Ghirardi, Rimini, and Weber) \cite{GRW}, and Continuous Spontaneous Localization, CSL \cite{CSL}. Indeed, there are now also relativistic versions of those theories, 
for instance see \cite{Tumulka2006, Bedingham2011, Rel-Pearle, Bed-D}, and proposals to have such spontaneous collapse tied to aspects of gravity have been considered by L. Diosi \cite{DIOSI} \& R. Penrose \cite{Penrose}.  For  a recent review  see  \cite{granGB}.

In the following, we briefly describe the CSL version for the setting of non-relativistic single-particle quantum mechanics. The theory is defined by two equations: 
\begin{itemize}
 \item [i)] A modified Schr\"odinger equation, whose solution is: 
\end{itemize} 
\begin{equation}\label{csl1}
 \ket{\psi, t}_{w}= \hat{\mathcal{T}}\exp{\left(-\int_{0}^{t}dt' \left[i\hat{H}+\frac{1}{4\lambda}[w(t')-2\lambda \hat{A}]^{2}\right] \right)} \ket{\psi,0}_w,
 \end{equation}

where $\hat {\cal T}$ is the time-ordering operator, $\hat A$ is the ``collapse" operator, a hermitian operator in the corresponding Hilbert space, which determines towards which states does the collapse dynamics tend to drive arbitrary initial states, $\lambda$ is the collapse rate. Furthermore, $w(t)$ is a random classical function of time, of white noise type, whose probability is given by the second equation, 
\begin{itemize}
 \item[ii)] The probability rule:
\end{itemize}
\begin{equation}\label{CSL2}
 PDw(t)\equiv{}_w \langle\psi,t|\psi,t\rangle_w \prod_{t_{i}=0}^{t}\frac{dw(t_{i})}{\sqrt{ 2\pi\lambda/dt}}.
\end{equation}

For non-relativistic quantum mechanics $\hat A$ is a smeared version of the position operator: 
 \begin{equation} 
 \hat A = N \int dx e^{(x-y)^2/r_c^2} |x\rangle \frac{x+y}{2}\langle y| 
 \end{equation}
where $r_c $ is the smearing scale considered to be of the order of $r_c \sim 10^{-5} \text{cm}$). On the other hand, $\lambda$ must clearly be small enough to avoid conflict with tests of quantum mechanics, and big enough to result in a rapid localization of ``macroscopic objects". GRW theories suggested range: $\lambda \sim 10^{-17} \text{sec}^{-1}$. More recent considerations indicate that $\lambda$ is likely not a universal constant, but depends on the particle's mass \cite{Mass-dependence}. The proposals are being actively tested in various experimental settings which have led to important constraints on the relevant parameters (See\cite{Bassi}).

These modified stochastic dynamics can account for the breakdown of symmetries, so we consider incorporating those into ``inflationary cosmology" to deal with the above-mentioned shortcomings of the standard accounts. 
 
 \smallskip

 As we have seen, the current versions of the theory are described in terms of the modified space-time evolution of quantum systems, so we need to consider settings where space-time notions are already present and it makes sense to rely on them. That is, even if one thinks that space-time is ultimately emergent from a deeper theory, we must take classical space-time as given, when working with spontaneous collapse theories. In short, we will treat space-time in classical terms in the following \footnote{As noted, we do not really have an alternative, as the existing proposals for quantum gravity theories are not yet developed to the point of providing general workable frameworks. On the other hand, the usual treatments based on quantizing the metric perturbations seem rather problematic at the conceptual level: For instance,  the standard microcausality conditions demanding field operators to commute at spacelike separation, can only be implemented in terms of the causal structure of the background metric, while the complete metric includes, besides that, the contribution from the perturbations. See for instance page 384 of \cite{Wald} for more on this point.}.

\section{Semi-classical gravity}
Semi-classical gravity is a theory in which space-time is treated classically and the scalar field, as well as the rest of the matter, is treated using Quantum Field Theory (QFT) in curved space-time. At this point, we should note that various arguments have been put forward against this option, including a famous ``experiment'' discussed in \cite{page} involving precisely the exploration of the gravitational field (or the space-time metric) associated with a massive object placed in a quantum superposition of two localizations \footnote{Other objections have been reported in\cite{SC-objections} which have been, however, questioned for instance in \cite{refutal-objections, refutal-objections2}.}. The analysis in \cite{page} concludes that: 
\begin{itemize}
\item[a)] If there are no collapses of the quantum state, the theory conflicts with experimental evidence.
 
\item[b)] If there are collapses of the quantum state the theory is inconsistent.
\end{itemize} 

The last statement is based on the observation that the kind of collapse required implies $\nabla^{a}\braket{T_{ab}}\not=0$, while the other side of Einstein's equations automatically satisfies $\nabla^a G_{ab}=0$ (Bianchi Identity). 
 
On the other hand, in \cite{Energy Conservation} we argued that all reasonable paths to deal with the \textbf{M problem} (P1 P2 \& P3) lead to the very same problem. Thus, Semi-classical gravity (together with any resolution of the \textbf{M problem}), {\bf cannot } be considered as a fundamental characterization of the interface of geometrical and quantum aspects of nature.

We, therefore,  will view semi-classical gravity together with suitably adapted spontaneous collapse theories as an approximate description of the limited range of validity. It is convenient at this point to consider a hydrodynamics analogy. As is well known, the Navier-Stocks equations provide an adequate description of fluid behavior in a broad set of circumstances, but they are clearly not truly fundamental equations (fluids are made of molecules and atoms, and these are made of elementary particles and none of those more fundamental degrees of freedom are described in the N-S equations) and, in fact, some breakdowns occur (for instance at the onset of turbulence).

We take our approach (combining a semi-classical description of gravity together with a QFT treatment of matter) as likely valid in regions with no collapse events, but acknowledge that departures would occur when spontaneous collapses are involved. In order to make things more explicit and precise, we will use the following definition characterizing the situations in which semi-classical GR together with QFT is taken to provide an appropriate description.
 
\medskip

 Formally speaking, we consider the so called \textit{Semi-classical Self-consistent Configuration} (SSC) (first introduced in \cite{A-Diez}).

\textbf{Definition:} The set $ g_{\mu\nu}(x),\hat{\varphi}(x), \hat{\pi}(x), {\cal H}, \vert \xi \rangle \in {\cal H}$
represents a 
SSC iff $\hat{\varphi}(x)$, $\hat{\pi}(x)$ and $ {\cal H}$ corresponds 
 to QFT in CS over the space-time  (as  described,  say, in \cite{Wald1994}) with metric $g_{\mu\nu}(x)$, and the state (called the physical state)
$\vert\xi\rangle$ in $ {\cal H}$ is such that:

\begin{equation}
G_{\mu\nu}[g(x)]=8\pi G\langle\xi\vert \hat{T}_{\mu\nu}[g(x),\hat{\varphi}(x),\hat{\pi}(x)]\vert\xi\rangle^{( Ren)}.
 \end{equation}
Note that this formalism involves a degree of self-reference, in that the QFT construction depends on the space-time metric, while the latter is connected to the former via the semi-classical Einstein equation. In fact, it can be considered as a general relativistic version of the Schr\"odinger-Newton system\cite{Sch-New}, which is also self-referential in nature.


\subsection{Incorporating of spontaneous collapse to semi-classical gravity framework }

It involves a change in the quantum state, which requires a change in the space-time metric, which in turn requires a change in the Hilbert space to which the state belongs. It should not be looked as {\it jumps in states} but {\it jumps} of the form: 

\begin{equation}
\text{SSC 1} \to \text{SSC 2}
\end{equation}
 
A scheme is needed to interpolate between or join SSC's. 
That requires gluing conditions: for space-times (along a collapse hypersurface) and transition of quantum states in the corresponding Hilbert spaces. This procedure involves delicate issues (renormalization of the expectation value energy-momentum tensor, well-posedness of the initial value formulation of semi-classical gravity, which has been investigated in\cite{Benito}). 

On the other hand, we should note that an extension of any collapse theory from the non-relativistic quantum mechanics settings to the relativistic quantum field theoretical one is highly nontrivial.

\section{Inflation with collapse theories and semi-classical gravity}

The zero mode of the inflation field, $\hat{\phi}_0$, is taken to start in a highly excited (and sharply peaked) state, while the space-dependent modes are in the Bunch-Davies vacuum (BD or adiabatic vacuum) state $|0 \rangle$. The quantum state of the scalar field and the space-time metric satisfy Einstein's semi-classical equations, 
\begin{equation} \label{semicl}
G_{\mu\nu}=8\pi G\langle\xi\vert \hat{T}_{\mu\nu}\vert\xi\rangle,
\end{equation}
under those conditions, one obtains essentially the standard behavior for the background (see \cite{diez}), 
\begin{equation}
 a(\eta) = \frac{-1}{\eta H_I}
 \end{equation} 
and  can  adjust  for  a  {\it  ``slow-roll'' } behavior  for $ \langle \hat {\phi}_0 \rangle$ in
 $ (-{\cal T}, \eta_f ), \eta_f < 0 $\cite{ls1}.

We concentrate next on the space-dependent $\vec k \not= 0$ modes. We have studied the case of the individual collapse of a single mode using the SSC formalism and a natural gluing recipe. More recently a generic collapse for general situations has been studied in \cite{Benito}. However here, we will rely on a {\it practical procedure} which gives equivalent results as the more rigorous formalism. The simplification involves relying on a single construction of the quantum field theory and allowing the states that result from collapses to be represented by elements of the original Hilbert space. The changes in the energy-momentum tensor that result from the collapses are then treated perturbatively. 

%
In the vacuum state, the operators $ \delta{\hat \phi}_k$, $\hat \pi_k$ are characterized by gaussian wave functions centered on $0$ with uncertainties $\Delta {\delta\phi_k }$ and
$\Delta{{\pi}_k}$, and $\Psi (\eta, x) =0, h_{ij } (\eta, x) =0$.

The collapse modifies the quantum state, and the expectation values of $\delta\hat{\phi}_k (\eta)$ and $\hat \pi_k (\eta)$.

We will for simplicity assume that the collapse occurs mode by mode and that it is described by an adapted version of collapse theories\footnote{That is, each mode of the inflaton field is treated as an independent degree of freedom, and its corresponding state is subject to a dynamical evolution given by equations \ref{csl1}, \ref{CSL2}. }. 
 Our universe would correspond to one specific realization of the stochastic functions, one for each $\vec k$.
 
First, we consider the scalar perturbations $\Psi(\eta, x)$, which characterize seeds of structure and are eventually imprinted in the CMB temperature fluctuations. The Fourier decomposition of the semi-classical Einstein's Equations give:
\begin{equation}
-k^2 \Psi(\eta,{\vec k}) = 
 \frac{4 \pi G \langle \hat{\phi}_0'(\eta) \rangle}{a} \langle \hat \pi({ \vec k},\eta)\rangle 
\end{equation}
Note that this expression is derived directly from (\ref{semicl}) using the assumption of negligible entanglement between $ \phi_{0}$ and $ \pi({ \vec k})$.
 The quantity containing the information about the outcomes of the stochastic processes are now the expectation values $\langle \hat \pi({ \vec k},\eta)\rangle$ (for the different $ \vec k $ values) evaluated in the state resulting from the modified quantum dynamics incorporating the stochastic collapse. 
 As these results are clearly unpredictable, the best one can do is to estimate the most likely value of the cumulative contributions from all modes $\vec k$ to the quantities of interest, 
 which, in the case of CMB observations of the scalar perturbations are the coefficients $\alpha _{lm} $ of the decomposition in spherical harmonics of the temperature variations $\frac{\delta T}{\bar T} (\theta, \varphi) $ with angle 
 in the celestial two-sphere. These temperature variations can be expressed in terms of the so-called Newtonian potential $\Psi$ evaluated on the intersection of the Last Scattering Surface with our past light cone. As shown in \cite{pss}, those quantities can be expressed as a sum of contributions
 from all modes, an expression which takes an analogous form as that corresponding to a random walk, for which the {\it most likely value} can be readily computed once the concrete adaptation of a collapse dynamics is provided. 
 
Collapse theories, as exhibited previously, contain various free parameters. From equation (\ref{csl1}) one can read two, the collapse rate $\lambda$ and the collapse operator $\hat{A}$
which, in the context of non-relativistic quantum mechanics, is taken to be the smeared position operator.
In the context of relativistic quantum field theory in curved space times, the appropriate collapse operator must clearly be something else. The question of what exactly it should be is not one for which we have at this time a general answer.
 In \cite{cps} two interesting cases were studied. In the first one, the collapse operator was chosen as the field $\hat{\phi}(\vec{x})$, while in the second, it was the conjugate 
 momentum operator of the field, $\hat{\pi}(\vec{x})$. The form for $\lambda$ in each case, is chosen in order to lead to a scale-invariant power spectrum (which is indicated by the observations  \cite{p1} (we will ignore at this point the slight departure from perfect scale invariance). 
\begin{equation}
\begin{split}\label{opscol}
 & \hat{\phi}(\vec{x}) \hspace{4pt} \text{as the collapse operator gives:} \hspace{4pt} \lambda=\tilde{\lambda}k,\\
& \hat{\pi}(\vec{x}) \hspace{4pt} \text{as the collapse operator gives:} \hspace{4pt} \lambda=\frac{\tilde{\lambda}}{k}.
\end{split}
\end{equation}

 We further note that the resulting forms are also appropriate from the dimensional analysis stand point, namely 
 both the above choices correspond to the quantity $\tilde{\lambda}$ having a dimension of inverse time, just as the standard parameter in the nonrelativistic versions of GRW/CSL.

We note that we could,  instead,  have chosen to take as the collapse operators either $( -\nabla^2)^{-1/2} \hat \pi ({\vec x})$ or $(-\nabla^2)^{1/2} \hat \phi ({\vec x})$ correspondingly.
Why is this the appropriate choice of collapse operator is not clear at this time. In fact, this is an instance of the broader general question, namely what is the universal form of the collapse operator that ought to appear in the more fundamental version of the theory? All we can say at this point is that this corresponds to a simple option that produces reasonable results. 

In the long run, one expects a general theory expressing, in all situations, from particle physics to cosmology, the exact form of the CSL-type of modification to the evolution of quantum states. 
We fully expect such theory would likely involve gravitation playing a fundamental role with the collapse dynamics likely involving a 
 local dependence on the curvature of spacetime (as discussed for instance in \cite{BHinfo2}, motivated in part by considerations expressed in \cite{d1,  DIOSI, Collapse-N-Sch3}).

With those assumptions, the resulting expression for the power spectrum is then: 
\begin{equation} 
P_S (k) \sim ( 1/k^3) (1/\epsilon) { (V/M_{Pl}^4)}{ \tilde \lambda} {\cal T}
\end{equation}
Taking the GUT scale for the inflation potential, and standard values for the slow-roll parameters leads to agreement with observation for: $\tilde\lambda \sim 10^{-5} \text{Mpc}^{-1} \approx 10^{-19} \text{sec}^{-1}$. We note that this value is not very different from GRW theory's suggested value. 

There are explorations involving alternative choices for the collapse operators, leading to rather different results (see\cite{Vanin, Gabrieles, TPSingh}).

\section{ The tensor modes}

\medskip
We now turn to consider the evolution for the tensor perturbations $ h_{ij}$ resulting from the semi-classical Einstein equation. Retaining only dominant terms, we find:
\begin{equation}
(\partial^2_0-\nabla^2)h_{ij}+2 \frac{\dot a}{a} \dot h_{ij} =16\pi G \langle (\partial_{i} \delta \phi)(\partial _{j}\delta \phi) \rangle_{Ren} ^{tr-tr}
\label{gw}
\end{equation}
 where the object in the RHS stands for the corresponding renormalized term in the energy-momentum tensor of the scalar field, and indicates $tr-tr$  the transverse trace-less part of the expression. According with our previous discussion we take vanishing initial conditions in \cite{tensor-Modes, tensor-Modes1} for $ h_{ij}$ as the initial situation was considered to be completely homogeneous and isotropic. 

We can already note that the source term is quadratic in the collapsing quantities. Passing to a Fourier decomposition, we solve the equation: 
\begin{equation}
\ddot {\tilde h}_{ij} (\vec k, \eta)+2 \frac{\dot a}{a}\dot {\tilde h}_{ij} (\vec k, \eta) + k^2 {\tilde h}_{ij} (\vec k, \eta) =S_{ij} (\vec k, \eta) ,
\end{equation}
with zero initial data, and source term: 
\begin{equation}
S_{ij} (\vec k, \eta) =16\pi G \int \frac{d^3x}{\sqrt{(2\pi)^{3}}} e^{i\vec k \cdot \vec x} \langle (\partial_{i} \delta\phi)(\partial_{j}\delta\phi) \rangle_{Ren}^{tr-tr} (\eta, \vec x).
\end{equation}
The result is formally divergent, however, it seems clear we must introduce a cut-off as modes with sufficiently short wave-length can be expected to become suppressed by diffusive physics. Thus, we take the cutoff to correspond to the standard 
({\it scale of diffusion, Silk dumping with $p_{UV} \approx 0.078 \text{Mpc}^{-1}$}). 
\medskip

 Then the prediction for the power spectrum of tensor perturbations is:
 \begin{equation} 
P_{h} (k) \sim ( 1/k^3) { (V/M_{Pl}^4)^2} ({ \tilde\lambda}^2 {\cal T}^4 p_{UV}^5 / k^3),
\end{equation}
where ${\cal T}$ is the conformal time at the start of inflation taken for standard inflationary parameters as $10^8$ Mpc, while the power spectrum for the scalar perturbations is 
 
\begin{equation} 
 P_S (k) \sim ( 1/k^3) (1/\epsilon) { (V/M_{Pl}^4)}{ \tilde \lambda} {\cal T}
 \end{equation}
 
That is a very different relation between them than the corresponding ``standard one''. Tensor modes are, thus, not expected at the level they are being looked for. The expression for the tensor perturbations indicates that the 
sensitivity of our detection mechanisms must be improved by various orders of magnitude before such detection would be feasible. Also it seems that an improved outlook for such detection would be attained by looking at very large scales. 
 
We also have considered a simpler collapse model, and again obtained a {\bf reduced tensor mode amplitude} but with a different shape. 
These results exemplify how the considerations of conceptual nature, that motivated the work along the present lines, lead not only to a more transparent account for the emergence of cosmic structure, but to a prediction that differs substantially from that emerging from the standard treatment, and which is, for an important number of specific models, in better agreement with the empirical evidence than the later.

\section{The Eternal Inflation Problem}

One of the pressing issues afflicting most inflationary models is the problem of eternal inflation. This arises from the tendency of simple inflationary scenarios to create conditions where inflation continues indefinitely in certain regions of the universe, making the end of this period highly improbable.
The essence of the argument according to \cite{Linde}: Inflation is driven by the zero mode of the inflaton field, $\phi_0 (\eta)$, which sets into a ``slow roll condition". In a characteristic time, its displacement $\Delta_{Class} \phi$ slowly decreases the effective cosmological constant, $\Lambda_{eff}$, corresponding to the value of the inflaton's potential. However, it is argued that the inflaton is also subject to ``quantum fluctuations'' $\Delta_{Quant} \phi$, and one must determine whether or not the latter dominates over the classical displacement. The situations correspond to the following two conditions:
\begin{equation}\label{sinp}
|\Delta_{quan} \phi_{0}| \leq \Delta \phi_{0}
\end{equation}
\begin{equation}\label{updown}
|\Delta_{quan} \phi_{0}| \geq \Delta \phi_{0}
\end{equation}
 
 According to the usual arguments, in the first case the situation would not be problematic, but in the latter case, in some regions, there would be an increase of the inflaton's potential, and thus, of $\Lambda_{eff}$ ( in comparison with the mean), and the opposite in other regions. The former regions would then grow faster than the latter, and then, at slightly later times, they will represent a larger portion of the universe. In time, the regions where the stochastic fluctuations were mostly upward, would represent the overwhelming portion of the universe, so we are likely to find ourselves in one such region. The expectation is therefore that, in practice, inflation will never end.

This argument, as it stands, raises some serious ``concerns" :

\begin{itemize}
 \item [1)] It relies once again on an identification of quantum uncertainties with stochastic fluctuations. In fact, as noted previously, the quantum fluctuations by themselves do not indicate something is randomly changing in space or in time. We emphasize that the meaning of quantum fluctuations is purely associated with uncertainties of some quantities. 
 
 \item[2)] Moreover, as we have noted, inflation is driven by the inflaton's zero mode, which is, by definition, homogeneous and isotropic, so nothing pertaining to it could be taken as indicating that something happens in some regions, and something else occurs in others. 
\end{itemize}

Thus, as it stands at this point, there seems to be no solid argument for {\bf eternal inflation}. However, as we have seen, we should supplement the standard story in order to have a sensible account for the formation of structure, and when that is provided by a spontaneous collapse theory, we then do have actual stochastic fluctuations in the evolution of the inflaton field, in the notation it can be read as $\Delta_{quan}\phi=\Delta_{Stoch}\phi$.
Therefore, in this framework the issue must be faced, but from a rather different perspective: Compare the classical displacement $\Delta_{Class} \phi$ in a characteristic time, with the corresponding stochastic mean displacement, $\Delta_{Stoch} \phi$ in the same time, due to the effect of the spontaneous collapses. 

As we have shown, in CSL theories\cite{cps} the election of the collapse operator is open, and in \cite{cps} two scenarios have been studied (\ref{opscol}). 
Focusing now on the implementation where the field is taken as the collapse operator \footnote{We found the case in which the collapse operator is taken to be the momentum operator to be problematic due to the divergence encountered when dealing with the zero mode of the inflaton field ( i.e. $\lambda = \tilde{\lambda}/k$). This scenario was analyzed in \cite{LeonET}.}, with $\lambda=\tilde{\lambda}k $, it becomes clear that the zero mode is NOT subject to collapses, and so it seems that in this context we would not be facing the eternal Inflation problem. 
However, we must confront the fact that there are modes, with such a long wavelength that they would be {\it effectively acting} as the zero mode: the modes with wavelength larger than the particle (causal) horizon i.e. modes with $ k< k_{c} $. 
Thus, we must compare the stochastic fluctuations, 
\[ (\Delta_{Stoch} \phi )^2 = \int_0^{k_c} d^3 k \overline{\braket{\delta_k \phi}^2 }\]

with $ (\Delta_{Class} \phi )^2$ for all times during the inflationary period.
 \medskip
 
In the work \cite{EI} we considered a slightly generalized form for the {$k$} dependence of the collapse rate, namely

\begin{equation}\label{newprop}
 \lambda=\tilde{\lambda}\left(\frac{k^{\alpha + 1}}{(b+k)^{\alpha}}\right),
\end{equation}

which, in order to reproduce the satisfactory results found in\cite{cps}, must reduce to $\lambda=\tilde{\lambda}k$ in the limit $k\gg b$ for the modes that are visible in the CMB (and BAO): $ 10^{-3} Mpc^{-1} <k< 10^{2} Mpc^{-1}$. In the other limit, $b\gg k$, (\ref{newprop}) also accounts for the modes that could play a role in eternal inflation, the form of the collapse rate is:
\begin{equation}
 \lambda=\tilde{\lambda}\left(\frac{k^{\alpha+1}}{b^\alpha}\right).
\end{equation}
 
 \medskip
 
The analysis indicated  the  condition for no-eternal inflation, namely,

\[
\frac{\Delta_{Stoch}}{\Delta_{class}}<1,
\]

  defines a region of parameter space where the \textbf{Eternal inflation} problem is avoided. This  is marked in yellow in the following graph: 
 \medskip

\begin{figure}[h!]
\includegraphics[scale=.60]{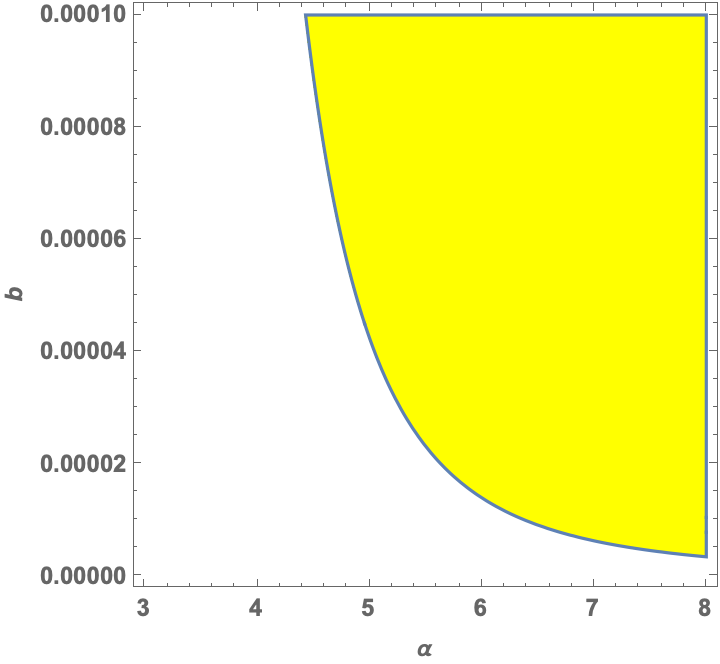}
\centering
\caption{In the last plot, the units are in $\text{Mpc}^{-1}$, and the yellow region represents where there is NO eternal inflation problem. The $\alpha$ scale was chosen arbitrarily while $b$ was selected to retain compatibility with CMB and BAO bounds.}
\end{figure}

 $b$ is in units of $10^{-5} \text{Mpc}^{-1}$. The freedom present in the modified quantum dynamics provided by spontaneous collapse theories opens the path for avoiding problems that appear to afflict many specific models of inflation, particularly the simplest ones.
 
 \section{Conclusions}
 
The cosmological setting is one where our current ideas require the use of both quantum theory and general relativity to account for the empirical observations that have recently become plentiful and increasingly precise. There is, however,  a general lack of corresponding precision in the specifications of the theoretical and conceptual frameworks employed in many of the analysis. Particularly problematic is the manner in which certain quantum concepts are used, often neglecting subtle difficulties relating to quantum theory itself. We have noted that this is the case, specifically concerning the notion of ``quantum fluctuations", where the apparently innocuous step of passing from viewing them as true quantum uncertainties and actual stochastic variations, allows one to bypass questions like how and when the cosmological homogeneity and isotropy of the initial situation depicted in the inflationary account, led to a universe teaming with inhomogeneities and anisotropies (necessary for the emergence of starts, planets, life and observers). 

We have presented an approach where the issue is addressed explicitly, relying on a version of quantum theory that resolves the measurement problem, which at the same time provides a reasonable account of the relevant periods of our universe's evolution, and of the empirical data behind our cosmological explorations. In fact, as we have seen, the approach we have been following leads to rather different expectations regarding the observability of primordial gravity waves produced during inflation (and which are in better accord with the current empirical situation), as well as a path to avoid the problem of eternal inflation.

The approach also offers paths to address other outstanding problems in the interface between gravitation and quantum theory. In particular, we have shown detailed studies presenting a resolution of the {\it black hole information loss puzzle} \cite{BHinfo, BHinfo2, BHinfo3, BHinfo4, BHinfo5, BHinfo6} (for a discussion of contrasting views  see\cite{BHinfo7}), and schematic analysis of how the proposal could serve to address {\it the problem of time in canonical approaches to Quantum Gravity} \cite{Benefits}. 


We conclude that focusing on the conceptual difficulties afflicting quantum theory in the form of the measurement and related problems, and, in particular, by adopting the path represented by the spontaneous objective collapse proposals, leads to rather satisfactory results concerning problematic aspects of inflationary cosmology as well as other outstanding problems in present-day physics.
There is, of course, much work to be done before we can consider these as definite answers, and it should be stressed that we are probably just at the beginning of the required exploration.

\section{Acknowledgments}

\noindent This manuscript was prepared as an invited paper for the Lemaître 2024 Conference, Vatican Observatory, Castel Gandolfo, Italy, June 17-21 2024. This work was partially supported by UNAM-DGAPA-PAPIIT grants IG-100124 and IN-101724 (RLLS), and Conhacyt grant 140630.

\section{ Appendix: Illustrating the Measurement Problem}

This serious shortcoming of the standard version of quantum theory is surprisingly ignored by large segments of the physics community including people working in cosmology, which,  as noted by Bell (see \cite{QMforCosmologists}), should be one of the groups most concerned with the issue. The measurement problem can be connected directly to the fact that the standard ``textbook"  version of quantum theory involves two different evolution processes for the states of a quantum system: the unitary Schr\"odinger evolution (or $U$ process in Penrose's terminology) describing the autonomous dynamics of the system, and the reduction or collapse of the state (or $R$ process in Penrose's terminology), resulting from what is called a measurement. When viewing the complete system, including the system of interest, the measuring device, and even the observer, as subject to a quantum description (accepting that quantum theory is universally valid), the shortcomings become apparent.  That is, the theory contains no general, precise, and clear rules indicating what amounts to {\it a measurement} and thus, fails to specify when each of the two different evolution rules applies. 

There is, of course, by now a vast literature on the subject exemplified by\cite{MP}, and this is not the place to offer a broad review on the matter. However, a simple example serves to illustrate some of its crucial aspects.

Consider a characterization of the setting for an experiment in which all aspects of the situation are exhaustively described. Namely, one is provided with the system's complete Hamiltonian, including the part describing the detection apparatuses as well as the quantum state of the system of interest, and that of the measuring devices. One then expects the theory to tell us what are the possible outcomes of the experiment, as well as the corresponding probabilities. Failure in this regard should clearly serve as an indication that we face at least some level of incompleteness.

So, let us consider a situation involving a single non-relativistic free particle which, for simplicity, we restrict to $1+1$ dimensions. Let the particle be described by a highly localized wave packet that we take to be a Gaussian centered at $x=0$ with a position uncertainty $\Delta$. That is the particle's wave function (in position space) is $\psi(x,t=0)=Ne^{-x^2/ \Delta^2}$.
 Take the particle-free Hamiltonian to be $H_0=\hat{P}^2/2m $. Now take two detectors labeled $ 1 \& 2$ to be simple two-level detectors, located at $ x=D$ and $x=-D$, with states $\lbrace |- \rangle_i , |+ \rangle_i \rbrace$ for $ i= 1, 2$ respectively, with the first corresponding to the unexcited detector state and the second to the excited detector. The detector's free Hamiltonian $H_i=\epsilon |+ \rangle_i 
{}_i\langle +|$. Finally, we take each particle-detector interaction Hamiltonian to be $ H_{int-i}= g\delta (x\pm D) [ |+ \rangle_i 
{}_i\langle -| + |- \rangle_i 
{}_i\langle +| ]$ . Let the initial state of the whole system correspond to the particle in the above state together with the two detectors in the unexcited state, namely

\begin{equation} \label{fuin}
|\Psi(t=0)\rangle = \int dx \psi(x, t=0) |x\rangle \otimes |- \rangle_1 \otimes |- \rangle_2
\end{equation}

We evolve this state with the full Hamiltonian $ H_T = H_0 + H_1 +H_2 + H_{int-1} + H_{int-2} $ and we find the complete state at time $ t$ is given by:

 \begin{equation} \label{decomp1}
 \begin{split}
|\Psi(t)\rangle &= \int dx \psi_(x,t)_N|x\rangle \otimes |- \rangle_1 \otimes |- \rangle_2 +
\int dx \psi(x,t)_L|x\rangle \otimes |+ \rangle_1 \otimes |- \rangle_2\\
&+ \int dx \psi(x,t)_R|x\rangle \otimes |- \rangle_1 \otimes |+ \rangle_2
+ \int dx \psi(x,t)_D|x\rangle \otimes |+\rangle_1 \otimes |+ \rangle_2
\end{split}
\end{equation}

which we readily interpret as indicating the amplitudes for non-detection (N), detection at the left detector (detector 1) (L), detection at the right detector (detector 2) (R), and detection at both the left and right detectors (i.e double detection) (D), given by the corresponding components.

At this point is worth emphasizing at least two problematic aspects of the presented situation, concerns that indeed arise from the standard quantum theory, \textbf{The unsatisfactory aspect of the Von Neumann approach}\footnote{Also known in the literature as measurement or macro-objectification problem \cite{VNew}.} and \textbf{the basis problem} \footnote{Indeed, these two problems were the main motivation for Ghirardi, Rimini, and Weber to postulate their first model of collapse theories, GRW \cite{Ghirardi:1985}. }.
\begin{itemize}

\item \textbf{The unsatisfactory aspect of the Von Neumann approach} becomes apparent in the previous situation when considering the question of how to interpret equation (\ref{decomp1}): the interaction between the detectors and the particle leads to a superposition of what we would like to consider as the possible outcomes: $N$, $L$, $R$, and $D$. However, there is nothing indicating that there are any outcomes at all. In the literature, this is referred to as the outcome aspect of the measurement problem, as after the 
evolution, the complete system's state does not represent a situation involving a definite outcome (in contrast with, let's say (\ref{fuin}, which shows a definite state for the detectors). Given that (\ref{decomp1}) represents an entangled state between the detectors and the particle, it is not possible to associate a specific state to the detector. As is stated in 
\cite{granGB} usually one attempts to deal with this evocating once more the {\it reduction process}. However, this type of analysis was supposed to account for such process and not to rely on it.
 This, in turn, reflects a serious shortcoming of standard Quantum Mechanics: The fact that it incorporates two different dynamical evolutions, $U$ and $R$ (in Penrose's words \cite{UyR}), without providing
 a clear rule indicating when each one is meant to apply, i.e. when is a process to be considered as a measurement? what is the quantity being measured? and exactly at what time is the measurement to be considered as having taken place? 
 The example shows explicitly that, even when the complete description of the system and measuring devices is given, that is, the quantum state of the whole system, as well as the free and interacting Hamiltonian, the theory lacks the elements to discern what are the possible outcomes of an experiment. 
\end{itemize}

\begin{itemize}
\item
\textbf{The basis problem} is a related concern illustrated by noting that we could have chosen to characterize the state of the detectors using a different basis, as for instance the following one:

 \begin{equation} 
|U\rangle = |- \rangle_1 \otimes |- \rangle_2
\end{equation}
 representing the fully ``Unexcited'' state, 
 \begin{equation} 
|S\rangle = \frac{1}{\sqrt{2}} ( |+ \rangle_1 \otimes |- \rangle_2 + |- \rangle_1 \otimes |+ \rangle_2
\end{equation}
 representing the `` Symmetrically excited'' state, 
 \begin{equation} 
|A \rangle = \frac{1}{\sqrt{2}} ( |+ \rangle_1 \otimes |- \rangle_2- |- \rangle_1 \otimes |+ \rangle_2
\end{equation} 
 representing the `` Anti-symmetrically excited'' state,
 \begin{equation} 
|D\rangle = |+ \rangle_1 \otimes |+\rangle_2
\end{equation}
 representing the `` doubly excited'' state,
whereby the state at time $t$ could be written as:

\begin{equation} \label{decomp2}
\begin{split}
|\Psi(t)\rangle& = \int dx \psi(x,t)_U | x\rangle \otimes |U\rangle +
\int dx \psi(x,t)_S |x\rangle \otimes |S \rangle \\
&+ \int dx \psi(x,t)_S |x\rangle \otimes |A\rangle 
+ \int dx \psi(x,t)_D|x\rangle \otimes | D \rangle
\end{split}
\end{equation} 
 which, following the same interpretational structure, would lead us to conclude that what we have are amplitudes and thus probabilities for finding the detectors in one of the various states $ |U\rangle , |S\rangle, |A\rangle, |D\rangle$. The first and last options are identical to the corresponding ones in the previous ``interpretation", but the second and third clearly represent something completely different (and which our experience would tell us is not what we find in an actual laboratory experiment).
 
 However, the point is that the theory, as it stands, without any further input (like that based on our intuitions) is unable to tell us which one of the two interpretational options is the appropriate one, i.e. which is the one that represents what actually happens. In short, the theory, as it stands, without an unequivocal clear interpretational framework involving something like a precisely stated and unambiguous reduction postulate, only indicates that the result of the evolution is the state in eq. (\ref{decomp1}) (or eq. (\ref{decomp2}) and does not indicate that we should rely on any specific decomposition of it \footnote{Furthermore, one cannot simply adopt the view that all decompositions are equally valid (ignoring the issue of what would that even mean), because, 
 as clearly shown by the Kochen-Specker theorem \cite{K-S}, it is impossible to consistently assign values to all physical quantities simultaneously.}.
 \end{itemize}

The example serves to illustrate an important aspect of the measurement problem, exhibiting also the fact that it becomes even more daunting in the context of applications to cosmology, where one cannot identify entities playing the role of observers or measuring apparatuses.


\newpage


\begin{thebibliography}{100}

\it \bibitem{Lemaitre} \it G. Lemaître, \it{``The Beginning of the World from the Point of View of Quantum Theory''} Nature {\bf 127}, 706 (1931).

\bibitem{lqg2} {\it C. Rovelli,} {\it ``Loop Quantum Gravity'', Living Reviews in Relativity}, {\bf 11}, 1–69 \it{(2008)}.

\bibitem{strig1} G. Horowitz, {\it ``String Theory as a Quantum Theory of Gravity''}, Proceedings of 12th Int. Conf. on General Relativity and Gravitation, edited by N. Ashby, D.F. Bartlett \& W. Wyss, 419–439 (1990).

\bibitem{cs1} L. Bombelli, J. Lee, D. Meyer, \& R. Sorkin, \it{``Space-time as a causal set''} {\it Phys.Rev.Lett.}, {\bf 59}, 521-524 (1987).

\bibitem{dt2} J. Ambj\o{}rn \& R. Loll, {\it ``Causal Dynamical Triangulations: Gateway to Nonperturbative Quantum Gravity''}, arXiv:2401.09399 (2024).


\bibitem{Naturalness} A. H. Guth, {\it ``The Inflationary Universe: A Possible Solution to the Horizon and Flatness Problems''}, Phys. Rev. D, {\bf 23}, 347–356 (1981).

 \bibitem{Norsen} T. Norsen, {\it ``Foundations of Quantum Mechanics: An Exploration of the Physical Meaning of Quantum Theory''}, Springer Cham, (2017).

\bibitem{UyR} R. Penrose, {\it ``The Emperor's New Mind: Concerning Computers, Minds, and the Laws of Physics''} Oxford Landmark Science, United Kingdom, (1989).
 


 \bibitem{EDM} O. Guerrero, L. Barr\'on-Palos \& D. Sudarsky, ``On the Quantum Uncertainty of the Neutron Electric Dipole Moment'' {\it Annals Phys.}, {\bf 469}, 169761 (2024).



\bibitem{Tim-trilema}
T.~Maudlin,
\newblock ``Three measurement problems'',
\newblock {\em Topoi}, 14(1):7--15, (1995).

\bibitem{SC-objections} Banks, T., Susskind, L., \& Peskin, \it{``Difficulties for the evolution of pure states into mixed states''}. Nucl. Phys., 44 B(234):125 M. E. (1984).

\bibitem{GRW} G.~C.~Ghirardi, A.~Rimini \& T.~Weber, 
``Quantum Probability \& Applications II'', L.~Accardi and W.~von Wandelfels eds., 
Lecture Notes in Mathematics, vol.~1136, Springer, Berlin (1985).

\bibitem{CSL} P.~Pearle, 
``Combining stochastic dynamical state-vector reduction with spontaneous localization'', 
Phys.\ Rev.\ A {\bf 39}, 5 (1989).

\bibitem{Bedingham2011} D.~J.~Bedingham, 
``Relativistic state reduction dynamics'', 
Found.\ Phys.\ {\bf 41}, 4 686-704, (2011).

\bibitem{Tumulka2006} R.~Tumulka, 
``A relativistic version of the Ghirardi–Rimini–Weber model'', 
J.\ Stat.\ Phys.\ {\bf 125}, 821-840 (2006).

\bibitem{Penrose} R. Penrose \&  C. J. Isham, \textit{``Quantum Concepts in Space and Time''},  Oxford University Press, (1986).

\bibitem{DIOSI} L.Diósi, ``A universal master equation for the gravitational violation of quantum mechanics,'' \textit{Phys. Lett. A}, \textbf{120}, 8, 377-381 (1987). 

\bibitem{d1} L.~Diosi, ``Gravitation and quantum mechanical localization of macro-objects'',
Phys.\ Lett.\ A {\bf 105}, 199 - 202 (1984). 

\bibitem{Bassi} S. Donadi, K. Piscicchia, C.Curceanu, et al., ``Underground test of gravity-related wave function collapse,'' \textit{Nature Phys.}, \textbf{17}, 74–78 (2021).

\bibitem{diez} 
A.~Diez \& D.~Sudarsky, 
``Towards a formal description of the collapse approach to the inflationary origin of the seeds of cosmic structure'', 
\textit{JCAP}, \textbf{045} (2012).

\bibitem{refutal-objections} 
Carlip, S.,``Is quantum gravity necessary?'' Class. Quant. Grav., 25:154010, (2008).

\bibitem{refutal-objections2} W. G. Unruh,  \& R. M. Wald, ``On evolution laws taking pure states to mixed states in quantum field theory'', Phys.Rev. D, 52:2176, (1995).
 
\bibitem{page} D.N. Page  \& C.D. Geilker, ``Indirect evidence for quantum gravity'', Phys. Rev. Lett. 47 (1981).

\bibitem{Energy Conservation} T. Maudlin, E. Ok\'on \& D. Sudarsky,  ``On the status of conservation laws in physics: Implications for semiclassical gravity", 
Stud. Hist. Philos. Sci.
69, (2020). 


\bibitem{Sch-New} M.~Bahrami, A.~Großardt, S.~Donadi  \& A.~Bassi, 

\bibitem{Linde}
A. D. Linde, ``Eternal chaotic inflation'', \textit{Mod. Phys. Lett. A} \textbf{1}, 81, (1986).
 
\bibitem{Vanin} J. Martin, V. Vennin \& P. Peter, ``Cosmological inflation and the quantum measurement problem'' {\it Phys. Rev. D}, {\bf 86}, 103524 (2012).



\bibitem{Gabrieles} G. Le\'on, G., \&  G. R. Bengochea,\textit{``Enlightening the CSL model landscape in inflation''}, \textit{Eur. Phys. J. C}, \textbf{81}(12), 1055 (2021).

\bibitem{TPSingh} 
S. Das, K. Lochan, S. Sahu, \& T. P. Singh, 
``Quantum to classical transition of inflationary perturbations: Continuous spontaneous localization as a possible mechanism'',
\textit{Phys. Rev. D}, vol. 88, no. 8, p. 085020, (2013).


\bibitem{EI} R.L. Lechuga \& D. Sudarsky, `` Eternal inflation and collapse theories", JCAP, Vol 01, 038 (2024).  

\bibitem{BHinfo}
 E. Ok\'on \& D. Sudarsky, ``The Black Hole Information Paradox and the Collapse of the Wave Function"
 {\it Found. Phys.} {\bf 45}, 461 (2015). 
 
 \bibitem{BHinfo2}
 S. K. Modak, L. Ortiz, I. Pe\~na \& D. Sudarsky, `` Non-Paradoxical Loss of Information in Black Hole Evaporation in Collapse Theories"
 {\it Phys. Rev. D} {\bf 91}, 124009 (2015).
 
\bibitem{BHinfo3}
D. Bedingham, S. K. Modak, \& D. Sudarsky, ``Relativistic collapse dynamics and black hole information loss'', {\it Phys. Rev. D}, {\bf 94}, 045009 (2016).

\bibitem{BHinfo4}
E. Okón, \& D. Sudarsky, ``Black Holes, Information Loss and the Measurement Problem'', {\it Found. Phys.}, {\bf 47}, 120 (2017).

\bibitem{BHinfo5}
E. Okón, \& D. Sudarsky, ``Losing stuff down a black hole'', {\it Found. Phys.}, {\bf 48}, 411-428 (2018).

\bibitem{BHinfo6}
S. K. Modak, \& D. Sudarsky, ``Collapse of the wavefunction, the information paradox and backreaction'', {\it EPJC}, {\bf 78} no.7, 556 (2018).

\bibitem{BHinfo7}
A. Perez, \& D. Sudarsky, ``A dialog on the fate of information in black hole evaporation'', chapter in Special Topic Collection Celebrating Sir Roger Penrose's Nobel Prize, {\it AVS Quantum Science}, {\bf 4}(4), 045602 (2022).


\bibitem{VNew}
A. Bassi \& G. Ghirardi, ``A general argument against the universal validity of the superposition principle''. \textit{Physics Letters A}, 275(5-6), 373-381, (2000). 

\bibitem{Benefits} E. Ok\'on, \& D. Sudarsky, ``Benefits of Objective Collapse Models for Cosmology and
Quantum Gravity" {\it FoP} {\bf 44}, 114 (2014).


\bibitem{A-Diez} A. Diez-Tejedor \& D.S., ``Towards a formal description of the collapse approach to the inflationary origin of the seeds of cosmic structure'', JCAP. 045, 1207, (2012) arXiv:1108.4928 [gr-qc].

\bibitem{Wald} R. Wald,  General Relativity, Chicago Univ. Pr., Chicago, USA,(1984).


\bibitem{Bed-D} D.J. Bedingham, ``Collapse models, relativity, and discrete spacetime'', Fundam. Theor. Phys. 198 (2020) 191; 
D.J. Bedingham, Relativistic state reduction dynamics, Found. Phys. 41 (2011) 686 [arXiv:1003.2774] 


 \bibitem{Tumulka} R. Tumulka, ``On spontaneous wave function collapse and quantum field theory'', Proc. Roy. Soc. Lond. A 462 (2006) 1897 [quant-ph/0508230] 
 
 \bibitem{Rel-Pearle} P. Pearle, ``Relativistic dynamical collapse model'', Phys. Rev. D 91 (2015) 105012 [arXiv:1412.6723].

 \bibitem{D-E1} E. Ok\'on \& D. Sudarsky, ``Less Decoherence and More Coherence in Quantum Gravity, Inflationary Cosmology and Elsewhere”, Found. Phys., 46(7), 852-879, (2016).

 
 \bibitem{decoh2} 
Stephen L. Adler, ``Why decoherence has not solved the measurement problem: A response to P.W. Anderson'', Studies in History and Philosophy of Science Part B - Studies in History and Philosophy of Modern Physics, 34(1), 135-142, (2003).


\bibitem{Mass-dependence}
P. Pearle and E. Squires, ``Bound state excitation, nucleon decay experiments and models of wave function collapse,'' \textit{Phys. Rev. Lett.} \textbf{73}, 1--5 (1994). 

\bibitem{Benito} B.A. Ju\'arez-Aubry, B.S. Kay, T. Miramontes, \& D. Sudarsky, On the initial value problem for semiclassical gravity without and with quantum state, collapses, JCAP 01 (2023) [arXiv:2205.11671].

\bibitem{tensor-Modes} G. Le\'on Garc\'{\i}a, A. Majhi, E. Ok\'on, \& D. Sudarsky, ``Reassessing the link between B-modes and inflation", {\it PRD }{\bf 96}, 101301(R) (2017).

\bibitem{tensor-Modes1} G. Le\'on
Garc\'{\i}a, A. Majhi, E. Ok\'on, \& D. Sudarsky,
``Expectation of primordial gravity waves generated during inflation''  Phys. Rev. D {\it PRD }{\bf 98} 023512 (2018).

\bibitem{QMforCosmologists} J.S. Bell \& A. Aspect, ``Quantum mechanics for cosmologists'', in \textit{Speakable and Unspeakable in Quantum Mechanics: Collected Papers on Quantum Philosophy}, Cambridge University Press, pp. 117-138,  (2004).

\bibitem{MP}
Norsen, T., \textit{Foundations of Quantum Mechanics: An Exploration of the Physical Meaning of Quantum Theory}, Springer Cham, (2017). 


\bibitem{LeonET}
G. León, ``Eternal inflation and the quantum birth of cosmic structure,'' \textit{Eur. Phys. J. C} \textbf{77}, 705 (2017).

\bibitem{Ghirardi:1985}
Ghirardi, G. C., Rimini, A., \& Weber, T., \textit{Quantum Probability and Applications II}, L. Accardi and W. von Wandelfels (eds.), \textit{Lecture Notes in Mathematics}, \textbf{1136}, Springer, Berlin (1985).


\bibitem{p1} P.~A.~R.~Ade {\it et al.} [Planck Collaboration], ``Planck 2013 results. XXII. Constraints on inflation," arXiv:1303.5082 [astro-ph.CO] (2013).


\bibitem{f1} S.~W.~ Hawking, ``The development of irregularities in a single bubble inflationary Universe", Phys.~Lett.~ B {\bf 115}, 295 (1982).

\bibitem{f2} A.~A.~Starobinsky, ``Dynamics of phase transition in the new inflationary Universe scenario and generation of perturbations," Phys.~Lett.~B {\bf 117}, 175-178 (1982).

\bibitem{f3} A.~H.~Guth and S.~Y.~Pi, ``Fluctuations in the new inflationary Universe'', Phys.~Rev.~Lett. {\bf 49}, 1110-1113 (1982).


\bibitem{i2} A.~L.~Linde, "A new inflationary Universe scenario: a possible solution of the horizon, flatness, homogeneity, isotropy and primordial monopole problems", Phys.~Lett.~ B {\bf 108}, 389-393 (1982).


\bibitem{f4} J.~M.~Bardeen, P.~J.~Steinhardt and M.~S.~Turner ``Spontaneous creation of almost scale-free density perturbations in an inflationary Universe", Phys.~Rev.~D {\bf 28}, 679 (1983).

\bibitem{f5} V.~F.~Mukhanov, H.~A.~ Feldman and R.~H.~ Brandenberger, ``Theory of cosmological perturbations", Phys.~Rep. {\bf 215}, 203-333 (1992).

\bibitem{dec1} D.~Polarski and A.~A.~Starobinsky, ``Semiclassicality and decoherence of Cosmological perturbations", Class.~Quant.~Grav. {\bf 13}, 377 (1996).

\bibitem{dec2} C.~Kiefer ``Origin of Classical Structure from Inflation", Nucl. Phys. Proc. Suppl. {\bf 88}, 255  and references therein, (2000).

\bibitem{dec3} C.~Kiefer and D.~Polarski,
 ``Why do cosmological perturbations look classical to us?'',
 Adv.\ Sci.\ Lett.\ {\bf 2}, 164,
 [arXiv:0810.0087 [astro-ph]], (2009).

\bibitem{pss} A.~Perez, H.~Sahlmann and D.~Sudarsky,
 ``On the quantum origin of the seeds of cosmic structure,''
 Class.\ Quant.\ Grav.\ {\bf 23}, 2317,
 [gr-qc/0508100], (2006).

\bibitem{s} D.~Sudarsky,
 ``Shortcomings in the Understanding of Why Cosmological Perturbations Look Classical'',
 Int.\ J.\ Mod.\ Phys.\ D {\bf 20}, 509,
 [arXiv:0906.0315 [gr-qc]], (2011).

 
\bibitem{s2}  J. Berjon, E. Ok\'on, D. Sudarsky, ``Critical review of prevailing explanations for the emergence of classicality in cosmology '',  Phys. \ Rev.\ D . {\bf 103}, No 4, 043521, (2021).

 
\bibitem{ss2} D.~Sudarsky,
 ``The Inflationary Origin of the Seeds: quantum theory and the need for novel physics"
 General Relativity, Cosmology and Astrophysics Fundamental Theories of Physics Vol. {\bf 177}, 349 (2014).

 
\bibitem{ls1} G.~Leon and D.~Sudarsky,
 ``The Slow roll condition and the amplitude of the primordial spectrum of cosmic fluctuations: Contrasts and similarities of standard account and the 'collapse scheme',''
 Class. Quant. Grav. {\bf 27}, 225017 (2010).


\bibitem{p} P.~Pearle, ``Combining stochastic dynamical state-vector reduction with spontaneous localization,"
Phys.\ Rev. \ A {\bf 39}, 2277 (1989).

\bibitem{cps} P.~Canate, P.~Pearle and D.~Sudarsky,
 ``CSL Wave Function Collapse Model as a Mechanism for the Emergence of Cosmological Asymmetries in Inflation'', Phys. Rev. D, 87, 104024 (2013).
 
\bibitem{Wald1994} R.~M.~Wald 
``Quantum Field Theory in Curved Spacetime and Black Hole Thermodynamics'',
Chicago Lectures in Physics, Chicago: The University of Chicago Press,
ISBN 0-226-87027-8, (2013).

\bibitem{granGB}
Bassi, A., \& Ghirardi, G.,``Dynamical reduction models'', \textit{Physics Reports}, 379(5), 257-426, (2003).


\bibitem{Collapse-N-Sch3} R.~Penrose, ``On gravity's role in quantum state reduction'',
Gen. Relat. Gravit., {\bf28}, 581 (1996).



\bibitem{K-S} Kochen, S., and Specker, E. P. ``On the problem of hidden variables in quantum mechanics", J. Math. Mech., {\bf 17}, 59–87 (1967). 

\end{thebibliography}
\end{document}